\documentclass[a4paper]{ISOStyle}
\usepackage{epsfig}
\usepackage{graphicx,subfigure}

\def\eps@scaling{.95}
\def\epsscale#1{\gdef\eps@scaling{#1}}

\def\plotone#1{\centering \leavevmode
    \epsfxsize=\eps@scaling\columnwidth \epsfbox{#1}}

\def\plotfiddle#1#2#3#4#5#6#7{\centering \leavevmode
    \vbox to#2{\rule{0pt}{#2}}
    \includegraphics{#1}}

\begin{document}

\title{Far-Infrared Emission from Dust in the ISOPHOT Virgo Cluster 
Deep Sample}

\author{Richard J. Tuffs}
  \institute{
     Astrophysics Division, Max-Planck-Institut f\"ur Kernphysik,
     Saupfercheckweg 1, 69117 Heidelberg, Germany 
     Richard.Tuffs@mpi-hd.mpg.de}
\maketitle 

\begin{abstract}

We review the characteristics of the dust continuum emission from normal
galaxies, as revealed by the ISOPHOT Virgo Cluster Deep Survey 
(Tuffs et al. 2002; Popescu et al. 2002, Popescu \& Tuffs 2002b).
\end{abstract}

\section{Introduction}

The ISOPHOT Virgo Cluster Deep Survey (Tuffs et al. 2002; Popescu et al. 2002)
represents the {\it deepest survey} (both in luminosity and surface 
brightness terms) of normal galaxies yet measured in the Far-Infrared (FIR). 
The survey consists of 63 gas-rich Virgo Cluster galaxies selected from the 
Virgo Cluster Catalog (VCC; Binggeli, Sandage \& Tammann 1985; see also 
Binggeli, Popescu \& Tammann 1993) and measured using the ISOPHOT instrument 
(Lemke et al. 1996) on board ISO (Kessler et al. 1996). 

The fundamental incentive for
choosing the VCC as the basis of a statistical sample for ISOPHOT
was that a luminosity- {\it and} volume - limited sample
of cluster periphery and cluster core galaxies representative of the field and
cluster environments, respectively, could be observed down to the least 
luminous dwarf galaxies reachable with ISOPHOT. This should
allow an investigation of the \newline strength and time-dependence of all 
manifestations of star formation activity and its relation to intrinsic 
galaxy properties such as Hubble type or sheer overall size. From an observational point of view the Virgo cluster has the advantage that it
is situated at high galactic latitude and is close to the ideal distance for
the detection of dwarf galaxies with ISOPHOT.
The VCC has also a full
representation of morphological types of normal gas rich galaxies, 
including  quiescent systems and even, to some extent, low surface 
brightness objects, ranging from bright ($B_{\rm T}\,\sim\,10$) 
giant spirals down to blue compact dwarfs (BCDs) and irregular galaxies 
at the completeness level of $B_{\rm T}\,\sim\,18$. Thus, the VCC is ideal for
providing the basis for statistical investigations of the FIR properties of gas 
rich galaxies in the local universe spanning a broad range in star-formation 
activity and morphological types, including dwarf systems.

\begin{figure}[htb]
\includegraphics[scale=0.5]{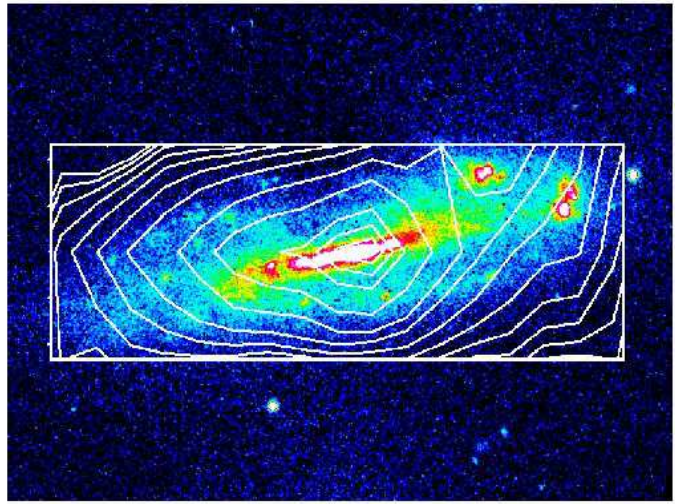}
\caption{The Virgo Cluter spiral galaxy VCC~66. Contour plots of the dust 
emission observed with ISOPHOT at 100\,${\mu}$m are overlaid on the R-band
image.} 
\end{figure}

\section{The observations}

The observations were done using the C100 and C200 detectors, in ISOPHOT's  
``P32'' observing mode (Tuffs \& Gabriel 2002a,b), which uses the focal plane 
chopper in conjunction with a spacecraft raster to rapidly sample
large areas of sky. The observing wavelengths were 60, 100 and 
170\,${\mu}$m. The ``P32'' mode allowed
the entire optical extent of each target down to the
25.5\,mag\,arcsec$^{-2}$ B-band isophote and adjacent
background to be scanned, while still maintaining a
spatial oversampling. This allowed both spatially integrated
FIR photometry as well as information on the morphology of the
galaxies in the FIR to be extracted. The data were reduced using the new
P32 software algorithm of Tuffs \& Gabriel (2002a,b). An example of an 
observation of the Sb spiral VCC~66 is given in Fig.~1.

From the 63 galaxies observed
(61 galaxies at all three FIR wavelengths and 2 galaxies 
only at 100 and 170\,${\mu}$m) we detected 54 galaxies at least at one 
wavelength and 40 galaxies at all three wavelengths. The averaged 
3\,$\sigma$ upper limits for integrated flux densities of 
point sources at 60, 100 and 170\,${\mu}$m are 43, 33 and 58\,mJy, 
respectively. The faintest 3\,$\sigma$ upper limits
at the three wavelengths are 30, 20 and 40\,mJy. 

\begin{figure*}[htb]
\plotfiddle{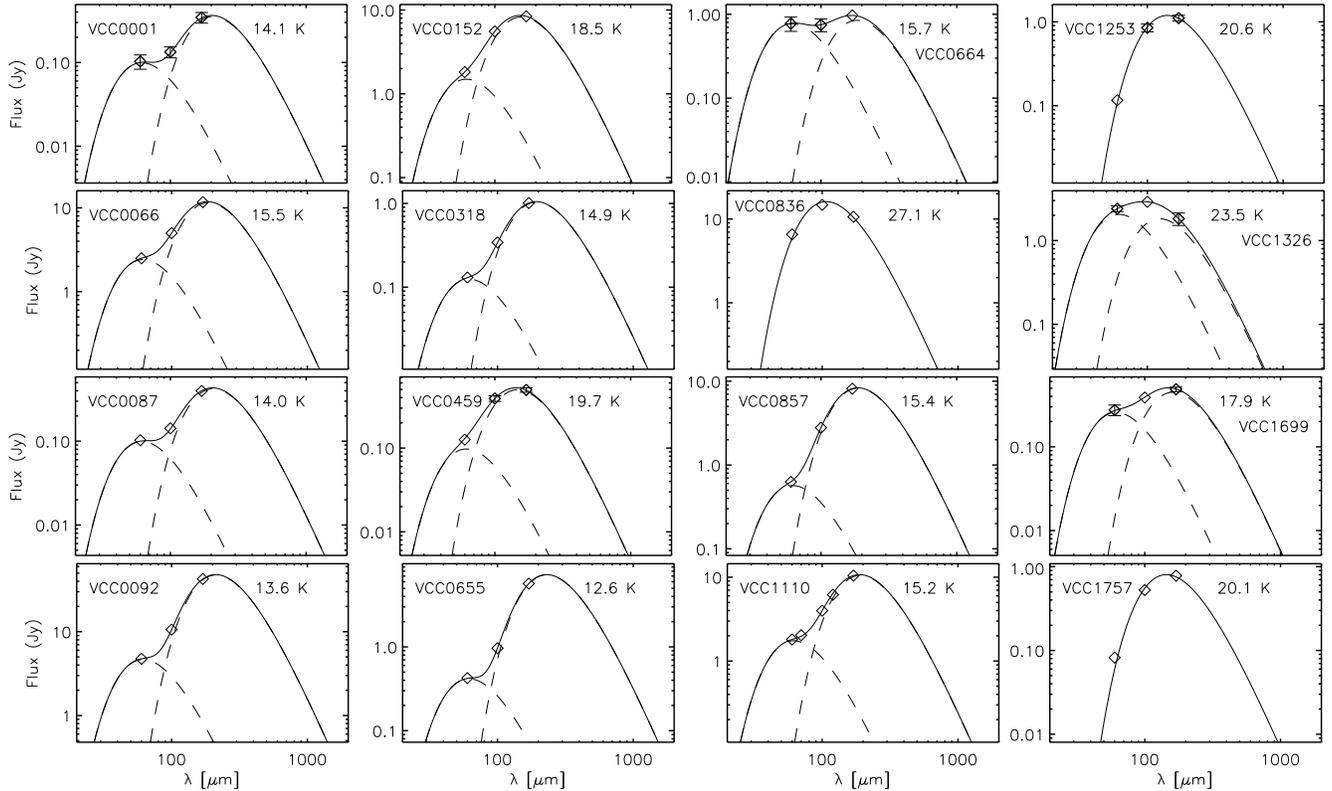}{4.5in}{0.}{95.}{95.}{-260.}{-380.}
\caption{Examples of FIR SEDs from the ISOPHOT Virgo Cluster Deep Sample
(Popescu et al. 2002). The colour corrected 
flux densities at 60, 100 and 170\,${\mu}$m are plotted together with their 
associated error bars. One galaxy, VCC~1110,has additional measurements at 
70 and 120\,${\mu}$m.
The two modified black-body functions which best fitted the data points are 
plotted with dashed-lines. The temperature of the warm component 
is constrained to be 47\,K. The fitted temperature of the cold component is 
marked near each fit. The sum of the two fitting functions is plotted as the
solid line.  Some galaxies (see text) don't show evidence for
two dust components and their SEDs are fitted with single component 
modified black-body functions, plotted as solid lines.}
\end{figure*}

The sample of VCC galaxies selected for ISOPHOT also formed a substantial part 
of samples observed with ISO using the ISOCAM MIR camera in pass bands 
centred at 6.9 and 15\,${\mu}$m (Boselli et al. 1997b, 1998) and the LWS at the
158\,$\mu$m [CII] fine structure gas cooling line 
(Leech et al. 1999).
In particular, the latter observations provide complementary information
about the energetics of the interstellar gas, the sources of [CII] emission
within the interstellar medium (Pierini et al. 1999, 2001), and
the role played by different stellar populations in the gas heating.

\section{The FIR spectral energy distributions}

We fitted the observed spectral energy distributions (SEDs) with a 
superposition of two modified black-body 
functions, physically identified 
with a localised warm dust emission component associated with HII 
regions (whose temperature was constrained to be 47\,K), and a 
diffuse emission component of cold dust. The two temperature components are
in fact predicted by the SED modelling of Popescu et al. (2000a), which
self-consistently analyses the UV/optical/NIR and the FIR/submm SEDs, and can
account for both the integrated flux densities and the surface brightness
distribution over the whole spectral range. The model\footnote{Full details of
the model are given by Popescu et al. (2000a), Misiriotis et al. (2001) and 
Popescu \& Tuffs (2002a).}, which includes  solving 
the radiative-transfer problem for a realistic distribution of
absorbers and emitters, considering realistic models for dust, taking into
account the grain-size distribution and stochastic heating of small grains and
the contribution of HII regions, has not only the power to derive the
star-formation rates and star-formation histories, but can also
predict the relative contribution of the young and old stellar populations to
the dust emission as a function of wavelength (see Fig.~3). Despite the
 complexity of the
model, which calculates a continuous distribution in dust temperatures, it 
can be seen that a natural outcome of this modelling technique is the 
prediction of a diffuse cold component of dust emission
powered by a combination of non-ionising UV and optical-NIR photons,
and a warm component of dust emission corresponding to the ensemble
of discrete HII regions. This corresponds to what we have seen 
in the ISOPHOT maps of Virgo galaxies. 

Thus, the emerging result from the FIR SED fits was that most of the Virgo
galaxies from our sample 
 require both warm and cold dust emission components to be fitted. The cold 
dust temperatures is broadly distributed, with a median of 18\,K
(Popescu et al. 2002), some $8-10$\,K lower than would have been predicted 
by IRAS. The corresponding dust masses were correspondingly found to be
increased by factors of typically  $6-13$ with respect with previous IRAS 
determinations. As a consequence,
the derived gas-to-dust ratios are much closer to the canonical value of$\sim
160$ for the Milky Way but with a 
broad distribution of values (Popescu et al. 2002).

\begin{figure}
\includegraphics[scale=0.50]{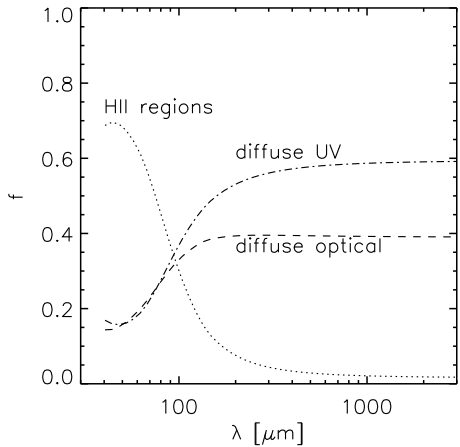}
\caption{The fractional contribution 
of the three stellar components to the FIR emission, for the case of NGC~891 
(Popescu et al. (2000a)}
\end{figure}

A good linear correlation is found between the ``warm FIR'' 
luminosities and the H${\alpha}$ equivalent widths (EW), supporting 
the assumptions of our constrained spectral energy distribution (SED) 
fit procedure. We also found a 
good non-linear correlation between the ``cold FIR'' luminosities
and the H${\alpha}$ EWs, consistent with the prediction of Popescu
et al. (2000a) that the FIR-submm emission should mainly be due to
diffuse non-ionising UV photons. Both the ``warm'' and the ``cold''
FIR luminosity components are non-linearly correlated with the 
(predominantly non-thermal) radio luminosities (Popescu et al. 2002). The 
validity of the FIR-radio correlation was thus tested using for the first 
time measurements of the bulk of the dust emission in quiescent normal 
galaxies, making this a highlight result of the ISOPHOT Virgo investigation.

Another highlight result is the calculation of the
percentage of stellar light re-radiated by dust. Previous estimates based on the
IRAS Bright Galaxy Sample (BGS; Soifer \& Neugebauer 1991)
have established a canonical value of 30\% for the
fraction of starlight to be re-radiated in the FIR in
the local universe. However, this value refers to relatively bright
FIR sources in which the bulk of the dust emission is radiated in
the IRAS 60 and 100\,${\mu}$m bands, and is not representative
of quiescent systems like the Virgo galaxies.
In addition it takes no account of measurements longwards of 120\,${\mu}$m, not
available at that time. The percentage of stellar light re-radiated by dust was
investigated by Xu \& Buat (1995), using an indirect estimate for the total FIR
luminosity. With the advantage of the new ISOPHOT data we calculate this 
percentage by using for the first time measurements of the bulk of the dust 
emission in quiescent normal galaxies.

By combining the luminosity of
dust emission with the observed UV/optical/NIR luminosities derived from
the data of Schr\"oder \& Visvanathan (1996), Boselli et al. (1997), Rifatto,
Longo \& Capaccioli (1995) and Deharveng et al. (1994) we show that the mean
percentage is $\sim 30\%$ for the later spirals in the 
ISOPHOT Virgo Cluster Deep Sample (Popescu \& Tuffs 2002b). 
This value is the same as the canonical value of 30$\%$ obtained for the 
IRAS BGS
by Soifer \& Neugebauer (1991). This is probably due to the fact that there 
are two factors influencing this percentage, and working in opposite 
directions. The addition 
of the ISO cold dust luminosity increases the FIR contribution to the total
bolometrics. But our sample consists of more quiescent galaxies than those
from  BGS and we expect them to have smaller FIR
contributions.
By the same token, it is probable that
the contribution of dust emission to the
total luminosity of the BGS galaxies will be greater than the 30$\%$
derived from IRAS.

\section{Trends with Hubble type}

Of particular interest are the results concerning the trends with Hubble type. 
A tendency was found for the temperatures of the cold dust component
to become colder, and for the cold dust surface densities 
(normalised to optical area) to increase with increasing lateness in the Hubble
type (Figs.~4a,b). A particularly surprising result was the low dust 
temperatures (ranging down to less than 10\,K) and large dust masses 
associated 
with the Virgo Im and Blue Compact Dwarf (BCD) galaxies. Another important 
trend is the increase of
the normalised (to K$^{\prime}$ band magnitude) FIR luminosity as we progress
 from the early to the later Hubble types (Fig.~4c). This result was
later confirmed by Bendo et al. (2002) for the RSA (Revised Shapley-Ames
Catalog) sample. A related result
was also obtained by Pierini et al. (1999) for the LWS data on 
Virgo galaxies, where a
strong correlation of normalised [CII] emission with H$\alpha$
equivalent widths was interpreted as a trend of increasing star-formation rate
along the Hubble sequence. 

\begin{figure*}[htb]
\plotfiddle{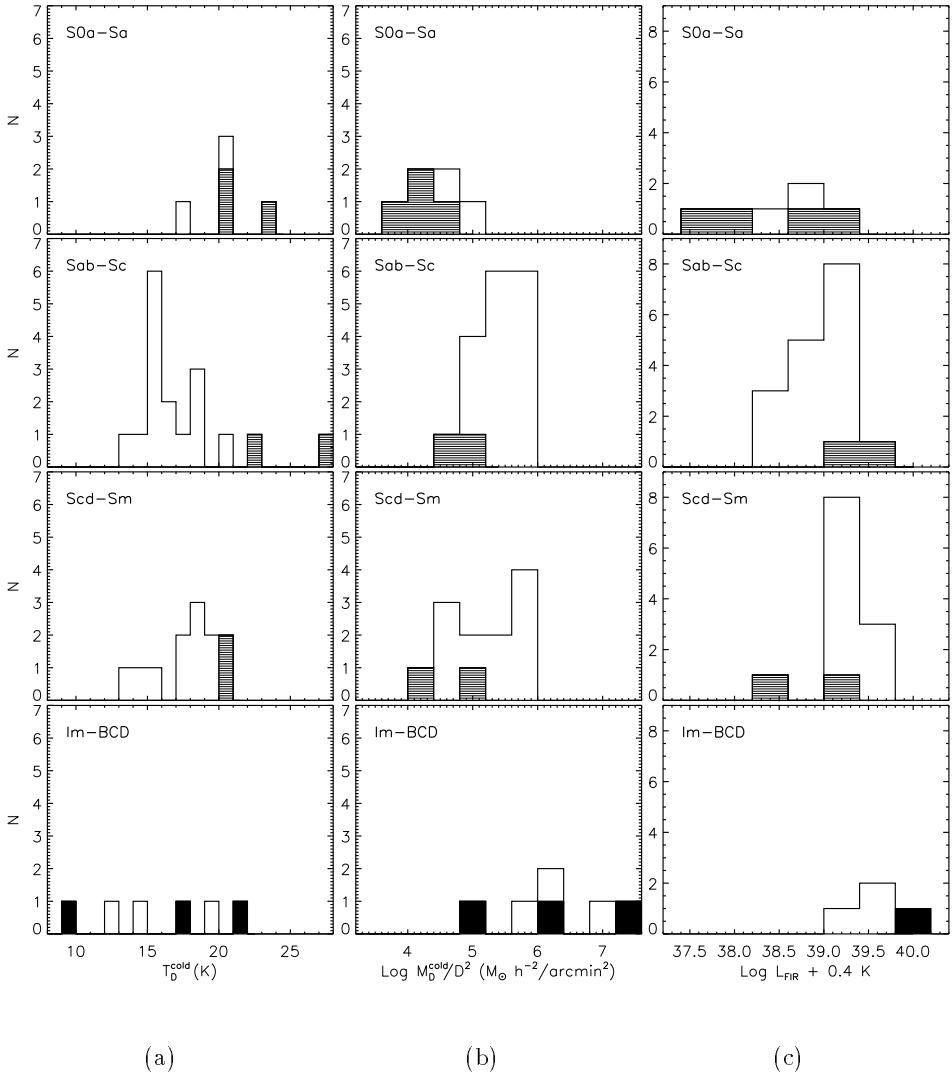}{5.5in}{0.}{95.}{95.}{-240.}{-300.}
\caption{Trends with Hubble type. The 
distribution of a) cold dust temperatures 
$T_{\rm D}^{\rm cold}$;
b) cold dust mass surface densities $M_{\rm D}^{\rm cold}/D^2$;
c) normalised FIR luminosity (to the K$^{\prime}$ band magnitude) for 
different Hubble types. The hatched histograms represent the distributions 
for the galaxies with SEDs fitted by only one dust component. The filled 
histograms represent the distributions for the galaxies with detections only 
at 100 and 170\,${\mu}$m. For the latter cases the dust temperatures are only 
upper limits and the dust masses are only lower limits.}
\end{figure*}

Finally, we found an
increase of the ratio of the dust emission to the total stellar emitted output
along the Hubble sequence. This correlation is quite strong, ranging from 
typical values of $\sim 15\%$ for early spirals to up to $\sim 50\%$ for some 
late spirals (Popescu \& Tuffs 2002b). This, together with the trend for a
decrease in the temperature of the cold dust, would suggest a trend
of increasing opacities with increasing star-formation activity. 
The extreme BCDs can have even higher 
percentages of their bolometric output re-radiated in the thermal infrared. 
This correlation can be also interpreted as a 
sequence from normal to dwarf gas rich galaxies, with the dwarfs having an 
increased contribution of the FIR output to the total bolometric output. These
findings could be important for our perception of the distant Universe, where,
according to the hierarchical galaxy formation scenarios, gas rich dwarf 
galaxies should prevail. We would then expect these galaxies to make a higher 
contribution to the total FIR output in the early Universe 
than previously expected. 
This, together with the cosmic-ray driven winds, in which grains can survive 
and be inserted in the surrounding intergalactic medium 
(Popescu et al. 2000b), could potentially change our view of the high 
redshifted Universe.

\section{Cold dust in the Virgo BCDs}

Perhaps the most intriguing result of this investigation 
are the masses and temperatures of the cold dust derived for
the Im and BCD galaxies in our sample. 
These systems are clearly differentiated from the spirals, having the 
highest dust mass surface densities (normalised to optical size), 
and the lowest dust temperatures. 
This is a particularly unexpected result,
since the IRAS observations of BCDs could be accounted for
in terms of dust heated locally in HII regions,
with temperatures of 30\,K or more.

The unexpected result that large amounts of cold dust exist in some
Virgo BCDs was interpreted by us as being indicative of dust 
surrounding the optical galaxy, originating in an external dust reservoir
fact, in two cases direct evidence was found of resolved emission at 170 
micron on scales of up to 10 kpc. 
To qualitatively account
for the FIR and optical extinction characteristics of BCDs, 
Popescu et al. (2002) proposed two scenarios invoking
collisionally or photon-heated emission from grains originating
in the surrounding intergalactic medium. 
In the one scenario, grains are swept up from a surrounding protogalactic
cloud and heated collisionally in an optically thin wind bubble blown
from the BCD. In the other, the grains are taken to be photon-heated
in an optically thick disk surrounding the optical galaxy. The disk is
indicative of a  massive gas/dust accreting phase which makes dwarf
galaxies sporadically bright optical-UV sources when viewed out of
the equatorial plane of the disk. In both scenarios the dust does not have a
galactic origin, but needs to exist in the immediate vicinity of the galaxies,
where it can either be heated by winds or can accrete into the dwarfs.

\end{document}